\newcommand{\sech}{\mbox{sech$\,$}}
\newcommand{\ts}{\textstyle}
\documentclass[showpacs,preprint,amsmath,amssymb]{revtex4}
\usepackage{graphicx}
 \begin{document}
 \title{ Shear Waves in an inhomogeneous strongly coupled dusty plasma}
 \author {M. S Janaki, D. Banerjee and N. Chakrabarti}
 \affiliation{ Saha Institute of Nuclear Physics,
 1/AF Bidhannagar Calcutta - 700 064, India.}
 \begin{abstract}

  The  properties of electrostatic transverse shear waves  propagating
  in a strongly coupled dusty plasma with an equilibrium density gradient
  are examined using the generalized hydrodynamic equation.  In the usual kinetic limit, the resulting equation has similarity to  zero energy Schrodinger's equation.
 This has helped in obtaining some exact eigenmode solutions in both cartesian and cylindrical geometries for certain nontrivial
density profiles. The corresponding velocity  profiles and the discrete eigenfrequencies are obtained for several interesting situations
  and their physics  discussed.

 \end{abstract}
 \maketitle
\section{Introduction}

In a  dusty plasma,  the interaction energy between neighbouring dust particles is often much larger than their thermal energy, leading to strong correlation effects between them.  This leads to the dust species to exist in a strongly coupled state
with a coupling parameter $\Gamma = (Z_d e)^2/aT_d > 1$, where $Z_de$, $T_d$
are the charge, temperature of the dust  grain and $a$ is the distance between
the dust grains. The propagation of collective modes in a plasma is influenced by the
strong correlation effects between dust particles and the dispersion relations
can be studied using the generalized hydrodynamics model\cite{kn:berk}-\cite{kn:kaw}
that is appropriate for studying wave motions in the long wave length limit.
For a strongly coupled
dusty plasma still existing in a fluid state, the existence of a novel transverse shear mode supported by solid-like rigidity of the medium has been predicted\cite{kn:kaw} and
also experimentally verified.  Molecular dynamics simulations carried out by Ohta and Hamaguchi\cite{kn:ohta} have identified such shear modes in the fluid
phase of Yukawa systems and obtained  dispersion relations that are in reasonably
good agreement with theoretical estimates.   In a medium with inhomogenous charge
distribution, the shear modes have been found to be driven unstable\cite{kn:sorasio} through the dynamics of dust charge fluctuations\cite{kn:am}. Another important outcome  arising out of the inhomogeneous effect is the coupling of the
shear and compressional modes that can exploited to find new mechanisms for excitation
of shear modes.  Such coupling also exists\cite{kn:yaro} among the longitudinal and
transverse dust lattice modes\cite{kn:nuno} and arises due to equilibrium dust charge gradients
and particle wake-interactions leading to a dust lattice mode instability.
In a non-Newtonian fluid, a coupling between inhomogeneous viscous
force and viscosity gradient leads to another novel instability of the
transverse shear mode\cite{kn:debabrata}. In a two dimensional complex plasma, viscoelastic vortical flows covering a wide scale
of spatial and temporal scales have been observed\cite{kn:ratyn} and the driving mechanism for such vortex formation is thought to be the charge inhomogeneity across such layer.

In the present work, we attempt to study the propagation of linear transverse shear
modes propagating in a strongly coupled dusty plasma having a density gradient
in the plane of propagation of the waves. The plasma is considered to be incompressible
so as to study the effects of density inhomogeneity on the dispersion characteristics of the shear modes in the absence of any coupling to the acoustic modes\cite{kn:am}.

\section{Basic Equations}
In a dusty plasma consisting of electrons, ions and dust particles, the charged dust component can easily go into the strongly coupled regime with large values of $\Gamma$ because of the large charge on the dust grains and/or small values of dust temperature
while the electrons and ions are in the weakly coupled regime.  In the regime
 $1<\Gamma<\Gamma_c$ ($\Gamma_c$ is the critical value beyond which  crystallization
sets in), the medium exhibits both viscous and elastic properties and is known
as a viscoelastic medium.  An intrinsic feature of viscoelastic
media is to exhibit the memory of accumulated shear strain showing viscous effects
on long time scales but elastic on short-time scales.  The elementary concepts of
viscosity and elasticity can be combined in a number of ways to obtain model equations
describing viscoelastic fluids.  Maxwell model of viscoelasticity\cite{kn:bird}  links the elastic part of stress to strain  via an exponentially decaying memory function.
A simple generalization\cite{kn:frenkel} of the Navier Stokes equation for an incompressible fluid taking into account Maxwell's relaxation elasticity consists in replacing the normal viscosity coefficient
$1/\eta$ by a viscoelastic operator $(1+\tau_m d/dt)/\eta$ leading to a generalized
momentum equation of hydrodynamics given by
\begin{equation}
    \left ( 1+ \tau_m   \frac{d}{d t}\right)
    \left[ \rho_d \frac{d{\bf u_d}}{dt}
    +\nabla p_d - Z_d e n_{d} {\bf E} \right ]
    =\eta \nabla^2 {\bf u}_d
    \label{gme}
    \end{equation}
where the convective derivative $d/dt \equiv \partial/\partial t + {\bf u}_d \cdot \nabla $.
    Here
    $\rho_d (=m_d n_d)$ , ${\bf u}_d$ and $Z_d e$ are the dust mass density, fluid velocity
    and charge respectively, $p_d$ is the dust pressure, $n_d$ is the number density,  $\bf E$ is the electric field,
    $\eta$  is the shear  coefficient of viscosity, $\tau_m$ is the viscoelastic relaxation
    time that accounts for memory effects.  The above equation takes into account
    the shearing elasticity of the  incompressible fluid, which though negligible for small
    values of $\eta$, becomes an important factor for large values of viscosity when
    solid like correlations begin to appear in the medium.
    A more general model applicable to a compressible fluid and involving both bulk and shear moduli  is obtained\cite{kn:berk},\cite{kn:kaw} by considering a nonlocal viscoelastic operator. 
For $\tau_m=0$, equation(\ref{gme}) reduces to the standard Navier-Stokes
hydrodynamic equation.  The generalized hydrodynamic model is known to support
propagation of undamped transverse shear modes in the kinetic limit characterized
by wave frequencies $\omega \gg 1/\tau_m $.  In order to study such modes, we take
the curl of the linearized  momentum equation (\ref{gme}) considering an
incompressible plasma ( $\nabla\cdot{\bf u}_d=0$), to obtain
\begin{equation}
 \tau_m   \frac{\partial^2}{\partial t^2}\nabla \times (\rho_d {{\bf u}_d})
   =\eta \nabla^2 (\nabla \times {\bf u}_d )
    \label{gme1}
\end{equation}
While deriving the above equation, electrostatic approximation has been assumed.
This assumption is justified for the low frequency modes being considered here.
For a two dimensional incompressible plasma, $\nabla \cdot {\bf u}_d=0$ implies
    the solution for the velocity as
    ${\bf u}_d = {\hat e}_z \times \nabla\psi(x,y)$, where $\psi(x,y)$ is velocity potential with $\nabla \times {\bf u}_d={\hat e}_z\nabla^2\psi$.  For an inhomogeneous plasma with $\rho_d = \rho_{0d} f({\bf r})$, we obtain from equation(\ref{gme1})
\begin{equation}
\nabla^4\psi+\gamma^2 f \nabla^2\psi + \gamma^2 \nabla f\times({\hat e}_z\times\nabla\psi)=0,
\label{gme2}\\
\end{equation}
where $\gamma^2 = {\omega^2}/{c_{sh}^2}\;$ ,$\;c_{sh}^2 ={\eta }/{\rho_{0d} \tau_m}$,
and $\rho_{0d}$ is the constant dust density.  An estimate \cite{kn:kaw} of $c_{sh}$  obtained
from one component plasma limit with the bulk viscosity coefficient neglected in
comparison to shear viscosity coefficient is given by
$$ c_{sh}^2 = 3 \frac{T_d}{m_d} \left(1-\gamma_d\mu_d+\frac{4}{15} u\right) $$
where $\gamma_d$ is the adiabatic index, $\mu_d$ is the compressibility and $u$
is the excess internal energy of the system given by $u(\Gamma)= -0.89\Gamma
+0.95\Gamma^{1/4}+0.19\Gamma^{-1/4}-0.81$.

For a uniform plasma, the governing equation describing the propagation of shear modes
is given by $\nabla^2(\nabla^2 \psi + \gamma^2 \psi)=0$
that gives the following dispersion relation
describing transverse shear modes:
\begin{equation}
\omega^2  = k^2 c_{sh}^2
\end{equation}
where $c_{sh}$ describes the propagation velocity of the transverse shear mode.

Equation(\ref{gme2}) can be solved for various equilibrium density profiles  to obtain the eigen-solutions and eigenvalues.  A nonlinear model for dust vortex
flows using the GH equation has been shown\cite{kn:shukla} to admit  several kinds
of vortex structures including the classical dipolar vortex solution.
It can be shown that for a special choice of equilibrium density profiles  of the following form
\[
f(r) = 1+\exp \left[\int \frac{J_1(\gamma r)}{J_1'(\gamma r)}d (\gamma r)\right ]
\]
(where $J$'s are Bessel functions of the first kind and the prime indicate derivative with respect to the arguments) the standard dipole vortex solutions $\psi(r,\theta)=J_1(\gamma r) cos\theta$
\cite{kn:nc} for the shear modes satisfying the equation
$\nabla^2\psi+\gamma^2 \psi=0$ can be recovered.

\section{Shear modes in an Inhomogeneous Plasma }

For a  plasma with nonuniform density, solutions of equation(\ref{gme2}) can be obtained by solving the following equation
\begin{equation}
(\nabla^2+\gamma^2 f)\nabla\psi = \nabla g
\label{gme3}
   \end{equation}
for appropriately chosen functional forms for the density profile $f$,
as well as $g$, where, for a uniform plasma, $\nabla g$ describes the
transformation to a moving frame of reference such that ${\bf u_d'} =
{\bf u_d}-{\hat e_z}\times \nabla g$, with $g$ satisfying the equation $\nabla^2 g=0$.
 Nontrivial solutions  for $g$ lead to an inhomogeneous term in equation (5) that is in general difficult to solve.
Throughout the present work, we shall consider $g=0$ and observe that equation(\ref{gme3})
has an analogy to Schrodinger's  equation with zero energy and the density
profile representing the potential.  We shall solve equation(\ref{gme3}) for a
few cases of specific interest.  The case of a uniform plasma is recovered
in the limit $f=1$.

(A) Firstly, we consider the propagation of shear waves in one dimension
with ${\bf u_d}=u_{y}(x) {\hat e_y}$ and density variation $f(x)= \sech^2(\alpha x)$
where $\alpha$ is the inverse of characteristic scale length of the density inhomogeneity.
 The chosen density profile is similar to a Gaussian profile that is
observed in plasmas mostly controlled by diffusion process and has the advantage that exact solutions can be obtained for such a profile
The eigen-values  of equation(\ref{gme3})
are obtained as
\begin{equation}
\gamma^2 = \frac{\omega^2}{c_{sh}^2} = n(n+1)\alpha^2
\end{equation}
where $n=1,2,3 \cdots$.  The corresponding eigenstates for $n=1,2$ \cite{kn:lekner} are given by

\begin{eqnarray}
u_{y}(x) &=& u_{y0} \tanh (\alpha x), \;\;\;\;\;\;\;\;\;\;\;\;\;\;\;\;\;\;\;\;\;\;\;\;\;\;\;\;\quad \frac{\omega^2}{c_{sh}^2} = 2\alpha^2 \nonumber \\
u_y(x)&=& u_{y0} \left[\sech^2 \left( \alpha x \right)-2 \tanh^2 \left (\alpha x \right )\right],\quad \frac{\omega^2}{c_{sh}^2} =6 \alpha^2
\end{eqnarray}
that correspond to localized solution.

(B)Next,  a two dimensional extension to case (A) can be obtained by considering
$ {\bf u_d} = u_y(x,y) {\hat e_y}$, with $f(x,y) = \sech^2(\alpha y)$.  Regular solutions to the inhomogeneous shear wave equation (\ref{gme3}) can be obtained \cite{kn:makowski} only if
the mode frequencies are discrete with $\gamma^2$  given by
\begin{equation}
\gamma^2_{ns} = \frac{\omega^2}{c_{sh}^2}={(s+n\pi)(s+n\pi+1)}\alpha^2,~~ {\rm{with}}~~s=0,1,2.. , n=1,2,3, \cdots
\end{equation}
The value of $\omega/c_{sh}$ is determined by the sum $s+n\pi$ with the lowest value
given by $\pi$ and the subsequent values given by $1+\pi,2+\pi,.$.  For each
such value of frequency, we get a corresponding shear mode eigenfunction.
The solutions for the velocity component $u_y(x,y) = -\partial \psi/\partial x $ are obtained as
\begin{equation}
u_y^{n,s}(x,y) = N_{ns}\sqrt{2 \alpha} \frac{\sin {n\pi \alpha x}}{(\cosh \alpha y)^{n\pi}} \; G_s^{\ts {n\pi+\frac{1}{2}}}\tanh \alpha y,
\end{equation}
where $ G_s^{n\pi+1/2}$ is a Gegenbauer polynomial, $N_{ns}$ is a normalization
constant.  The velocity profiles and the dispersion relations for a few cases are given below

\begin{eqnarray}
u_y^{1,0} &=& N_{10} \sqrt{2 \alpha} \sin {\pi \alpha x}
(\sech\alpha y)^\pi, \;\;\;\;\;\;
\quad \frac{\omega^2}{c_{sh}^2}={\pi(\pi+1)}\alpha^2\nonumber \\
u_y^{1,1} &=& N_{11}  \sqrt{2 \alpha} \left(\pi+\frac{1}{2}\right) \sin {\pi \alpha x}
(\sech \alpha y)^\pi \tanh \alpha y, \;\;\;\;\;\;
\quad \frac{\omega^2}{c_{sh}^2}={(1+\pi)(2+\pi)}\alpha^2 \nonumber \\
u_y^{1,2}&=&N_{12}\sqrt{2 \alpha} \left(\pi+\frac{1}{2}\right) \sin {\pi \alpha x}
(\sech \alpha y)^\pi \left[ 2\left(\pi+\frac{3}{2}\right)\tanh^2 \alpha y-1\right]
,
 \quad \frac{\omega^2}{c_{sh}^2}={(2+\pi)(3+\pi)}\alpha^2 \nonumber\\
\label{s2}
\end{eqnarray}
where
$$N_{10}^2 = \alpha \frac{\Gamma(\pi+1/2)}{\sqrt{\pi}\Gamma(\pi-1/2)},
N_{11}^2 = \frac{N_{10}^2}{2(\pi+1/2)}, N_{12}^2 = \frac{N_{10}^2}{2(\pi+1/2)(\pi+1)}$$
\begin{figure}[ht]
              \includegraphics[width=3.0in,height=3.0in]{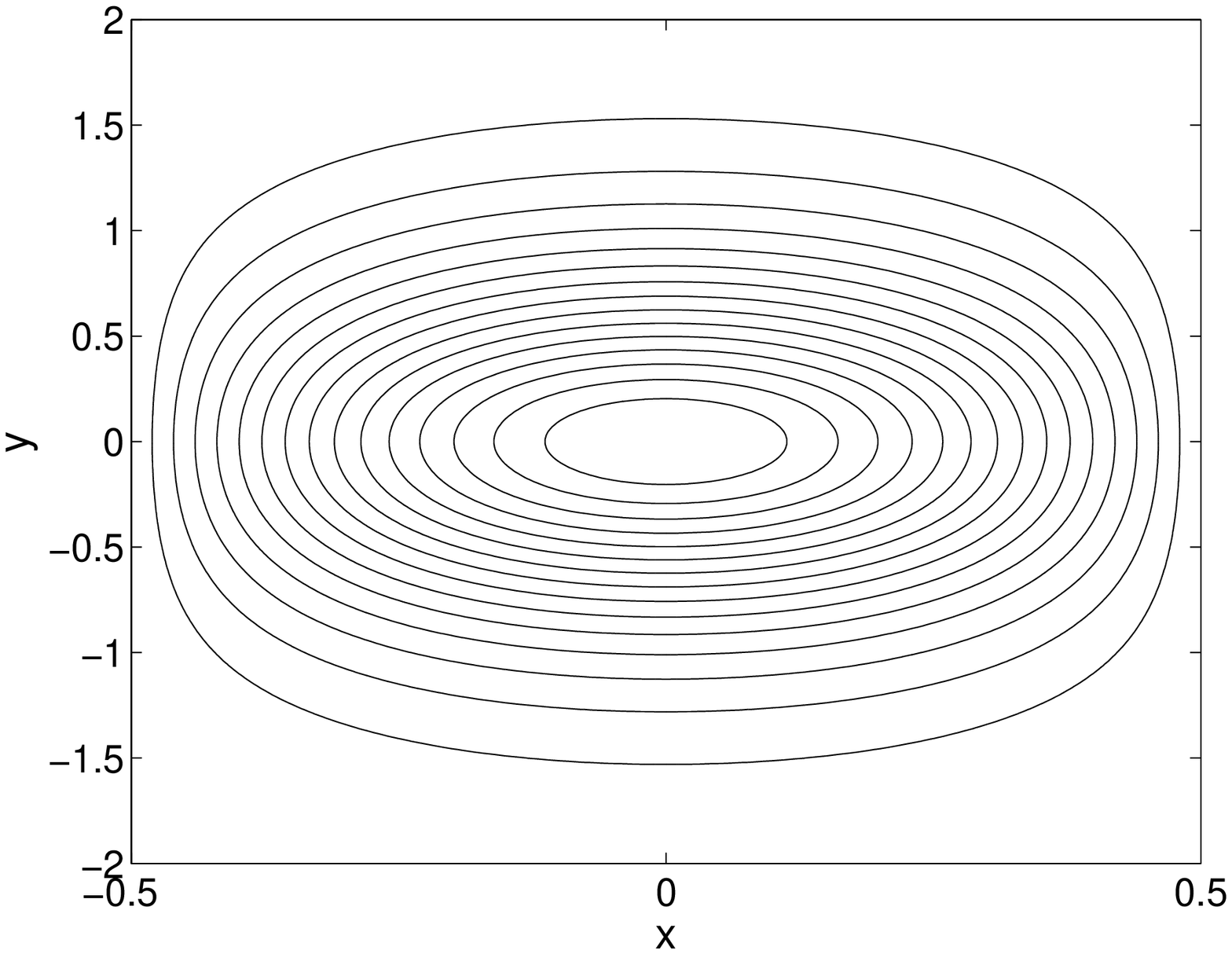}
              \caption{Velocity potential contours in the x-y plane
 for  $ s=0,n=1 $}
              \label{x1}
              \end{figure}
\begin{figure}[ht]
              \includegraphics[width=3.0in,height=3.0in]{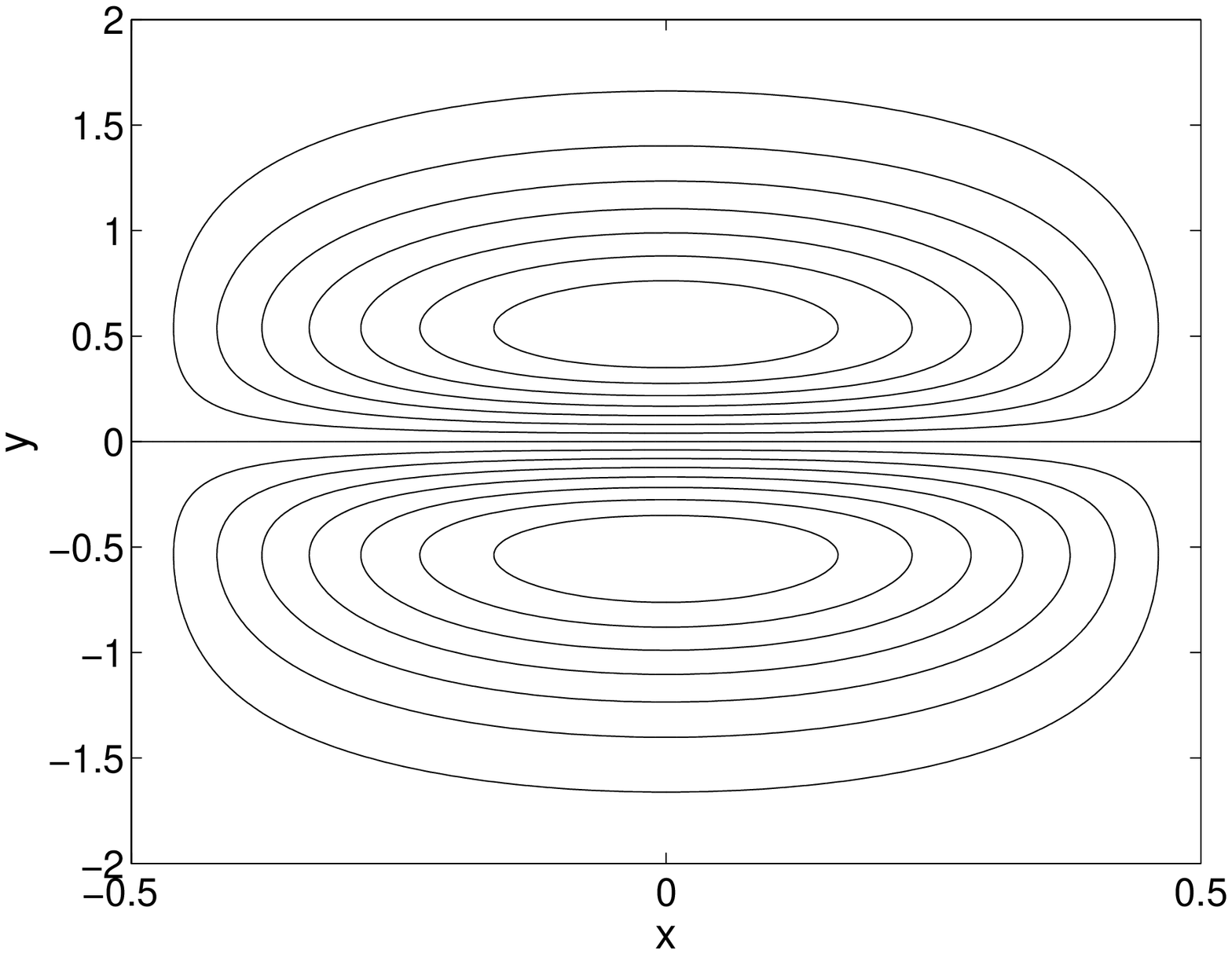}
              \caption{Velocity potential contours in the x-y plane for $s=1,n=1 $}
              \label{x2}
              \end{figure}
 \begin{figure}[ht]
              \includegraphics[width=3.0in,height=3.0in]{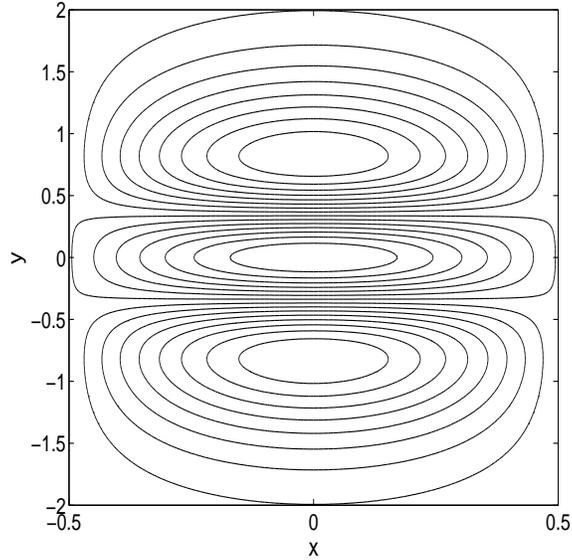}
              \caption{Velocity potential contours in the x-y plane for s=2,n=1}
              \label{x3}
              \end{figure}
Equation(\ref{s2}) shows that for a density gradient in the $y-$ direction, we obtain periodic solutions
in the x-direction, with localization in the y-direction.
The solutions corresponding to $n=1,s=0,1$ give rise to monopole and dipole
vortices respectively while the solution for $n=1, s=2$ represents a tripolar
vortex.  The corresponding velocity  potential contours are shown in figs.1-3.
The number of vortex structures in the figure grows as $s+1$ with the maximum value
of velocity potential lying at $x=0$ in each case.

(C)  Finally, we  consider a cylindrical plasma, with an equilibrium density profile
given by $f(r)=(1-\alpha^2 r^2)$.
In cylindrical coordinates, we get a coupled system
of two equations for the components $u_r$ and $u_\theta$ that are not easy to solve.
However, equation(\ref{gme3}) can be solved for the two scalars,
namely the cartesian components $u_j, (j=x,y)$ of the velocity vector ${\bf u}$,
as functions of $r,\theta$ with
the equation expressed in cylindrical coordinates with $z-$ symmetry. The components $u_r,u_\theta$ can be expressed in terms of $u_x,u_y$.  Making the substitution
$u_j(r,\theta)= \chi(r)\exp(im\theta)$ in equation(\ref{gme3}), we obtain
\begin{equation}
\frac{1}{r}\frac{\partial}{\partial r}\left(r\frac{\partial \chi}{\partial r}\right)
-\frac{m^2}{r^2}\chi+\gamma^2(1-\alpha^2 r^2)\chi = 0
\label{gme4}
\end{equation}
A physically well-behaved solution of equation(\ref{gme3}) is obtained in the form
\begin{equation}
u_j(r,\theta) = (\alpha r)^m e^{-\gamma \alpha r^2/2} {_1}F_{1}\left (\frac{1+m}{2}-
\frac{\gamma }{4 \alpha}, 1+m; {\gamma \alpha r^2 } \right ) \cos{m\theta}
\end{equation}
\begin{figure}[ht]
              \includegraphics[width=3.0in,height=3.0in]{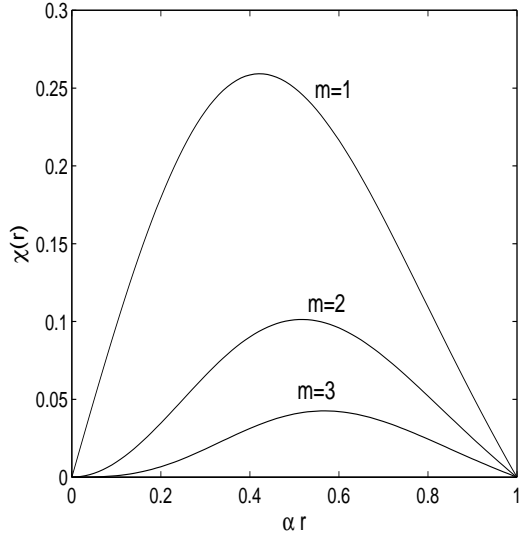}
              \caption{Eigenfunctions obtained for m=1,2,3 with $\omega/c_{sh}\alpha
= 4.62, 6.52, 8.42$ respectively.}
              \label{x3}
              \end{figure}
where $_1F_1(a,b,x)$ is a Kummer confluent hypergeometric function.  The physically acceptable solutions are obtained by considering that
$\chi(r) $ vanishes at  $r=0,r\alpha=1$.   This leads to the velocity profiles vanishing at the axis and at the boundary. These conditions give the eigenfrequencies
of the shear waves.  For each frequency, we obtain
two different types of eigenfunctions with different symmetries.
For the choice, $u_y=0$, we obtain the flow lines situated in the plane
$y=$constant and
the  components of ${\bf u}$ are obtained as
\begin{eqnarray}
u_r(r,\theta) = (\alpha r)^m e^{-\gamma \alpha r^2/2} {_1}F_{1}\left (\frac{1+m}{2}-
\frac{\gamma }{4 \alpha}, 1+m; {\gamma \alpha r^2 } \right ) \cos \theta \cos{m\theta}\nonumber \\
u_\theta(r,\theta) = (\alpha r)^m e^{-\gamma \alpha r^2/2} {_1}F_{1}\left (\frac{1+m}{2}-
\frac{\gamma }{4 \alpha}, 1+m; {\gamma \alpha r^2 } \right ) \sin \theta \cos{m\theta}
\end{eqnarray}
For the parabolic density profile chosen here, the radial part of eigenfunctions, $\chi(r)$ situated in the plane $y=constant$ are shown in fig. 4.
 The corresponding eigenfrequencies
$\omega/c_{sh} \alpha$ are given by $4.62$, $6.52$, $8.42$ for $m=1,2,3$ respectively.  For higher poloidal mode numbers, the peak of the
radial part of the eigenfunctions shifts towards the plasma edge.
A different choice ($u_x=0,u_y\neq 0$) will yield eigenmodes with the corresponding symmetry.
Various kinds of profiles in the $r-\theta$ plane
can be generated depending on the values of constants that determine
the functional form of the Kummer function.

\section{Summary}
In this work we have investigated the properties of shear waves in an inhomogeneous
plasma where gradients in plasma density are present. The equation governing the velocity potential is identical to zero energy Schrodinger's equation
with the density profile playing the role of potential.
The role of inhomogeneity is shown to give rise to a discrete set of allowed frequencies.
For a one dimensional
model, in presence of a density gradient in the direction of propagation, localized shear modes are obtained in contrast to the freely propagating modes with the
allowed mode frequencies  depending on the inhomogenity scale length.
  Shear modes
propagating in two dimensions with a density profile that is constant in the
x-direction and ${sech^2}$ type in the y-direction are shown to admit  multipolar vortex solutions.  In a cylindrical bounded geometry,  with a parabolic density
profile, exact analytical solutions for radial and azimuthal velocity components are
obtained in terms of Kummer hypergeometric functions.


\begin{thebibliography}{99}
\bibitem{kn:berk} M.S. Berkovsky, Phys. Lett. {\bf{A 166}},365 (1992).
\bibitem{kn:boon} J.P. Boon and S. Yip, Molecular Hydrodynamics, (Mc-Graw Hill, New York, 1980).
\bibitem{kn:kaw} P.K. Kaw and A. Sen, Phys. Plasmas, {\bf{5}},3552 (1998).
\bibitem{kn:ohta} H. Ohta and S. Hamaguchi, Phys. Rev. Lett. {\bf{84}},6026 (2000).
\bibitem{kn:sorasio}G. Sorasio , P. K. Shukla and D. P. Resendes, N. J. Phys.
{\bf{5}}, 81, 2003.
\bibitem{kn:am} A. Mishra, P.K. Kaw and A. Sen, Phys. Fluids, {\bf{7}}, 3188 (2000).
 \bibitem{kn:yaro} V. V. Yaroshenko, A. V. Ivlev, and G. E. Morfill, Phy. Rev. E
{\bf{71}},046405 (2005).
\bibitem{kn:nuno}S. Nunomura, S. Zhdanov, D. Samsonov, and G. Morfill, Phys. Rev. Lett. {\bf{94}},045001, 2005.
\bibitem{kn:debabrata} D. Banerjee, M.S. Janaki and N. Chakrabarti, N. Jour. Phys.
{\bf{12}}, 123031, 2010.
\bibitem{kn:ratyn} S. Ratynskaia, K. Rypdal, C. Knapek, S. Khrapak, A. V. Milovanov,
 A. Ivlev, J. J. Rasmussen, and G. E. Morfill, Phys. Rev. Lett. {\bf{96}},105010 (2006).
\bibitem{kn:bird} R. B. Bird, R. C. Armstrong, and O.
Hassager, Dynamics of Polymeric Liquids, (Wiley, New York, 1987, vol. 1).
\bibitem{kn:frenkel} J. Frenkel, Kinetic Theory of liquids, (Clarendon, Oxford, 1946).
\bibitem{kn:shukla} P.K. Shukla, Phys. Plasmas, {\bf{10}}, 1619 (2003).
\bibitem{kn:nc} M.S. Janaki and N. Chakrabarti, Phys. Plasmas {\bf{17}}053704(2010).
\bibitem{kn:lekner}   J. Lekner,  Am. J. Phys. {\bf{75}}, 1151 (2007).
\bibitem {kn:makowski} A.J. Makowski, Annals. Phys. {\bf{324}},2465 (2009).


\end{thebibliography}
\end{document}